\documentclass{article}

\PassOptionsToPackage{numbers, compress}{natbib}
\usepackage[preprint]{neurips_2026}
\usepackage[utf8]{inputenc} 
\usepackage[T1]{fontenc}    
\usepackage{hyperref}       
\usepackage{url}            
\usepackage{booktabs}       
\usepackage{amsfonts}       
\usepackage{nicefrac}       
\usepackage{microtype}      
\usepackage{xcolor}
\definecolor{linkblue}{RGB}{0, 102, 204}
\hypersetup{
    colorlinks=true,          
    linkcolor=linkblue,       
    citecolor=linkblue,       
    urlcolor=linkblue         
}

\usepackage{adjustbox}      
\usepackage{subcaption}

\usepackage{authblk} 
\usepackage{amsmath}
\usepackage{tikz}
\usetikzlibrary{shapes.geometric, arrows.meta, positioning}
\usepackage{units}
\usepackage{xspace}

\newcommand{\github}[1]{\href{https://github.com/#1}{\adjustbox{valign=c}{\includegraphics[width=8pt]{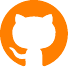}}}}

\title{FAIR Universe\\Weak Lensing ML Uncertainty Challenge:\\Dataset and Cosmological Inference Competition}

\author[1\textbf{*}]{\textbf{Biwei Dai}}
\author[2\textbf{*}]{\textbf{Po-Wen Chang}}
\author[3]{Erwan Allys}
\author[2]{Wahid Bhimji}
\author[4]{Boris Bolliet}
\author[4]{Thomas Borrett}
\author[2]{Paolo Calafiura}
\author[5,6]{Ragansu Chakkappai}
\author[7]{Yuan-Tang Chou}
\author[8]{No\'{e} Dia}
\author[2]{Sascha Diefenbacher}
\author[9,2]{Ibrahim Elsharkawy}
\author[2]{Steven Farrell}
\author[10]{Sacha Guerrini}
\author[6]{Isabelle Guyon}
\author[2]{Chris Harris}
\author[7]{Shih-Chieh Hsu}
\author[11]{Wassim Kablan}
\author[12]{Fran\c{c}ois Lanusse}
\author[4]{Celia Lecat}
\author[13]{Julia Linhart}
\author[14]{Yuezhou Ma}
\author[15,16]{Benjamin Nachman}
\author[17]{Shubhojit Naskar}
\author[4]{Andy Nilipour}
\author[8]{Laurence Perreault-Levasseur}
\author[3]{S\'{e}bastien Pierre}
\author[18,19]{Benjamin Remy}
\author[5,6]{David Rousseau}
\author[20,2]{Uro\v{s} Seljak}
\author[14]{Haonan Shangguan}
\author[8]{Sammy Nasser Sharief}
\author[14]{Yuanxu Sun}
\author[21]{Andreas Tersenov}
\author[6]{Ihsan Ullah}
\author[14]{Huikun Weng}
\author[4]{Licong Xu}
\author[8]{Justine Zeghal}
\author[7]{Yulei Zhang}
\author[14]{Hang Zhou}

\affil{Institute for Advanced Study,
\textsuperscript{2}Lawrence Berkeley National Laboratory,
\textsuperscript{3}\'{E}cole normale sup\'{e}rieure -- PSL,
\textsuperscript{4}University of Cambridge,
\textsuperscript{5}Universit\'e Paris-Saclay, CNRS/IN2P3, IJCLab,
\textsuperscript{6}ChaLearn,
\textsuperscript{7}University of Washington,
\textsuperscript{8}Universit\'{e} de Montr\'{e}al,
\textsuperscript{9}University of Toronto,
\textsuperscript{10}CEA Paris-Saclay,
\textsuperscript{11}Fran\c{c}ois Arago Centre, Astroparticule et Cosmologie (APC) laboratory,
\textsuperscript{12}Centre national de la recherche scientifique (CNRS),
\textsuperscript{13}New York University,
\textsuperscript{14}Tsinghua University,
\textsuperscript{15}SLAC National Accelerator Laboratory,
\textsuperscript{16}Stanford University,
\textsuperscript{17}Indian Institute of Technology Delhi,
\textsuperscript{18}University of Chicago,
\textsuperscript{19}NSF-Simons AI Institute for the Sky (SkAI),
\textsuperscript{20}University of California, Berkeley,
\textsuperscript{21}University of Crete
}
\affil[*]{\textbf{Equal contribution} (with remaining names in alphabetical order)}

\begin{document}
\maketitle

\begin{abstract}
Weak gravitational lensing is a powerful cosmological probe that enables direct mapping of the matter distribution and precise constraints on the fundamental parameters governing structure formation in the universe. In recent years, machine learning has shown strong potential for extracting cosmological information from weak-lensing data beyond traditional summary statistics. However, progress has been limited by the lack of standardized benchmarks that capture the complexity of the dataset, simulator imperfections, and various systematic biases encountered in realistic analyses. To address this gap, we present the FAIR Universe Weak Lensing Machine Learning Uncertainty Challenge, one of the NeurIPS 2025 competitions, as a benchmark dataset and evaluation study. Our dataset reflects realistic weak-lensing observations, incorporating multiple systematic biases, and thereby providing a controlled testbed for robust cosmological inference under uncertainties. We summarize the main methodological advances developed by the winners of our challenge, including data augmentation with conditional optimal transport, an AI-agent-driven approach for parameter inference, ensemble learning, Vision Transformer-based architectures, and physics-informed feature integration. By unifying the benchmark dataset and the winning solutions, this work provides a community resource for developing and evaluating machine learning methods for weak lensing cosmology and, more broadly, for scientific inference under systematic uncertainties. 
\end{abstract}


\section{Introduction}
Simulation-based inference (SBI) and related machine learning (ML)-based inference methods have become powerful tools across the sciences \cite{cranmer2020frontier}, enabling parameter estimation and model comparison in settings where the likelihood is intractable but forward simulations are available. Combined with modern machine learning, SBI can extract information from high-dimensional datasets that traditional summary statistics cannot capture, opening up analyses that were previously infeasible \cite{lanusse2023dawes}. However, its deployment in real scientific problems faces several persistent challenges: high-fidelity simulations are computationally expensive in many fields, restricting most realistic setups to a low training data regime \cite{modi2025sensitivity}; imperfect modeling of physical processes and measurement effects creates distribution shifts between simulations and real data that can bias inference \cite{villaescusa2021multifield}; and reliable quantification of uncertainties, which is essential for trustworthy scientific conclusions, remains an open methodological problem.

Cosmology, specifically the analysis of the large-scale structure of the universe, is a prototypical example of these challenges. Modern cosmological analyses rely heavily on simulations to interpret observations, and ML-based inference shows great promise in extracting the rich non-Gaussian information from the high-dimensional data produced by current and upcoming sky surveys \cite{fluri2018cosmological,Gupta2018a,dai2022translation,lu2023cosmological,dai2024multiscale,sharma2024comparative,cheng2025cosmological,jeffrey2025dark,von2025kids,zeghal2025simulation}, a task that traditional two-point analyses cannot accomplish. However, existing studies typically adopt different simulation setups and systematic treatments, making rigorous comparison difficult. The absence of a shared benchmark dataset further compounds the challenges outlined above, as it prevents the community from systematically assessing how different methods handle limited training data, distribution shifts, and uncertainty quantification.

To address these gaps, we present the first benchmark dataset for ML-based inference in weak gravitational lensing, one of the most informative cosmological probes of the matter distribution in the universe \citep{hoekstra2008a, kilbinger2015a}. The dataset, which underpins the FAIR Universe Weak Lensing ML Uncertainty Challenge, consists of simulated weak lensing maps that mimic the Hyper Suprime-Cam (HSC) Year-3 dataset \cite{Li:2021mvq}, with realistic systematic effects included to mimic real data analysis. Built around this dataset, the challenge is structured in two phases, targeting cosmological parameter inference with uncertainty quantification (Phase 1; one of the NeurIPS 2025 competitions) and out-of-distribution detection (Phase 2)~\cite{Dai:2026jme}; see our \href{https://github.com/FAIR-Universe/Cosmology_Challenge}{GitHub repository}~\github{FAIR-Universe/Cosmology_Challenge}. In this paper, we describe the dataset (publicly available on \href{https://zenodo.org/records/20056065}{Zenodo}), the Phase 1 task, evaluate the winning solutions, and present the final results from Phase 1, whose techniques we expect to advance both upcoming weak lensing analyses and the broader development of ML-based inference in science.

\section{Dataset}
\label{sec:dataset}
According to general relativity, the gravitational field of matter bends the path of light traveling through the universe. As light from distant galaxies traverses the inhomogeneous matter distribution between the galaxies and us, the galaxies' observed shapes become subtly and coherently distorted. This effect, known as weak gravitational lensing \cite{hoekstra2008a, kilbinger2015a}, can be measured statistically across many galaxies to reconstruct the projected matter distribution along the line of sight, typically expressed as a 2D convergence map. Weak lensing convergence maps directly encode information about cosmological parameters, making them one of the most powerful probes of cosmology.

Our dataset consists of simulated weak lensing convergence maps that mimic the data products of the HSC Year-3 survey \cite{Li:2021mvq,rau2023weak}, one of the leading current weak lensing experiments. Each map is generated by running cosmological simulations \cite{yu_feng_2018_1451799,Feng:2016yqz} and a ray-tracing algorithm \cite{jain2000ray, hilbert2009ray, Petri2016mocking}, under the HSC Y3 footprint, redshift distribution, and galaxy density. Each map has $1424 \times 176$ pixels. 

The dataset spans $101$ cosmological models in the $(\Omega_m, S_8)$ plane, where $\Omega_m$ is the total matter density of the universe and $S_8 = \sigma_8 (\Omega_m / 0.3)^{0.5}$ characterizes the amplitude of matter fluctuations. We additionally model various important sources of systematic uncertainty (e.g., baryonic feedback and photometric redshift uncertainty) in weak lensing analyses, which are parametrized by nuisance parameters that are sampled during data generation and must be marginalized over during inference. For each cosmological model, we generate $256$ maps with different random realizations and different draws of the nuisance parameters, yielding a total of $101 \times 256 = 25{,}856$ training maps. Each map carries a 5-dimensional label $(\Omega_m, S_8, T_{\mathrm{AGN}}, f_0, \Delta z)$, where the first two are the parameters of interest and the last three are nuisance parameters.
 
The training maps are released noiseless. Noise from intrinsic galaxy shapes can be added as a post-processing step during training. We adopt a noise level that matches the expected sensitivity of upcoming weak lensing surveys such as  \href{https://sci.esa.int/web/Euclid}{Euclid Space Telescope} and the \href{https://lsst.org}{Vera Rubin Observatory}. Full simulation details, systematics modeling, and noise specifications are provided in Appendix \ref{app:simulation}.

\section{Competition task and metric}
\label{sec:compeition}
Participants are tasked with developing ML models that can accurately determine point estimates of two cosmological parameters $\hat{\Omega}_m$ and $\hat{S}_8$ and quantify their one standard deviation uncertainties $\hat{\sigma}_{\Omega_m}, \hat{\sigma}_{S_8}$ from a suite of noisy weak lensing maps. Participants are provided with a labeled training dataset containing $101 \times 256$ convergence maps described in Sec.~\ref{sec:dataset}. Participants then submit their predictions to our Codabench website~\cite{Phase1_Codabench}. The model performance is ranked with the metric
\begin{equation}
    \text{score} = -\frac{1}{N_{\text {test}}} \sum_{i}^{N_{\text {test}}} \sum_{\boldsymbol{\Theta}_i} \left[ \frac{\left(\hat{\boldsymbol{\Theta}}_{i}-\boldsymbol{\Theta}_{i}^{\text {truth}}\right)^2}{\hat{\sigma}_{\boldsymbol{\Theta}_i}^2} + \log \left(\hat{\sigma}_{\boldsymbol{\Theta}_i}^2\right) + \lambda \left(\hat{\boldsymbol{\Theta}}_{i}-\boldsymbol{\Theta}_{i}^{\text {truth}}\right)^2  \right]~, \label{eq:inf_score}
\end{equation}
where $\boldsymbol{\Theta} = (\Omega_m, S_8)$. The first two terms correspond to the Kullback–Leibler divergence between the true posterior distribution and the Gaussian distribution with the predicted mean and standard deviation. We expect the posterior distribution to be pretty Gaussian and the correlation between $\Omega_m$ and $S_8$ to be small, thus the Gaussian approximation with a diagonal covariance matrix is reasonable. The third term is an MSE loss with weight $\lambda=10^3$ to penalize bad point estimates.\footnote{The MSE penalty term in Eq.~(\ref{eq:inf_score}) can further be normalized with the prior ranges of $\Omega_{m}$ and $S_8$ to minimize the difference between the MSE scales of the two parameters of interest.}

\section{Evaluation}
\label{sec:evaluation}

\paragraph{Public phase}
During the public phase of the Phase 1 competition, participants submitted their predictions of cosmological parameters and their uncertainties using the public test dataset we provided. The dataset contains $4{,}000$ realizations of weak lensing maps with ground truth cosmological parameters randomly sampled from the $101$ cosmological models described in Sec.~\ref{sec:dataset}. The submissions were then evaluated using Eq.~\ref{eq:inf_score} and ranked on the public leaderboard.

\paragraph{Final evaluation}
The leading public submissions were further evaluated on two distinct holdout sets: (A) $6{,}000$ realizations from the simulations with the original $101$ cosmological models in the training data, and (B) $4{,}020$ realizations from the simulations with $10$ cosmological models that were not seen in the training data. The holdout set (A) rewards performance on the original cosmology grid, whereas the holdout set (B) probes interpolation and generalization to unseen cosmological models. All holdout data were generated using the same pipeline code and priors as the training data. The scores obtained from the two holdout sets and the average of both together determine three final leaderboards to evaluate the leading public submissions for this competition. 

\paragraph{Bootstrapping robustness test}
To assess whether the final leaderboards could be affected by statistical fluctuations, we generated $10{,}000$ bootstrap samples and examined the score distributions for each participant. In each bootstrap sample, we sampled $4{,}000$ predictions with replacement from participants' predictions of both holdout sets (A) and (B). We then estimate the mean scores and the probabilities of achieving each rank using all bootstrap samples. We find that the bootstrapped mean scores and rankings are consistent with the final scores reported in Table \ref{tab:phase1_final_leaderboards}.

\section{Limitations}
\label{sec:limitations}

Our original benchmark is intentionally simplified to make the competition accessible to a broad community. In particular, we restrict the released data to a single redshift bin and a single HSC subfield, and we provide reconstructed convergence maps rather than the full galaxy catalog. We also model only a subset of the systematic effects relevant for a full weak-lensing analysis. These choices reduce the size of the released training set to below 10\,GB and lower the computational barrier to entry, but they also mean that the benchmark does not yet capture the full complexity of survey-grade analyses, including additional systematics such as intrinsic alignments, point spread functions, etc. 

Second, the public test set for the public leaderboard on Codabench contains different realizations of the same $101$ cosmological models that appear in the public training set. As a result, repeated use of the public leaderboard for model development may increase the risk of overfitting to this discrete set of cosmological models, even though the nuisance parameters and noise realizations vary. In hindsight, we could caution participants about the overlap of cosmological models in the public datasets to help mitigate the issue. Our final evaluation takes this issue into account by separately assessing performance on different leaderboards. 

Finally, the holdout set (B) only contains $10$ unseen cosmological models, so small differences between closely performing submissions on (B) should not be over-interpreted. 

To address some of these issues, we have generated a new training set with a higher effective number of independent samples (released on \href{https://zenodo.org/records/20056065}{Zenodo}), and a new larger test set that was sampled from $82$ cosmological models not present in the training set (will be released after the completion of Phase 2 competition to prevent test set leakage into the Phase 2 competition).
We plan for a series of data releases towards a permanent community benchmark. Further releases will increase the dataset size, include more realistic systematic effects, and cover larger sky areas.

\section{Results and best solutions}
\label{sec:results}

Table~\ref{tab:phase1_final_leaderboards} shows the final leaderboards evaluated on different holdout sets. On the leaderboard of the holdout set (A), the top three participants are \texttt{Cmbagent}, \texttt{Transatlantic Dream Team}, and \texttt{Shubhojit}. On the leaderboard (B), \texttt{Shubhojit} achieves the best score and is clearly separated from the rest, while the remaining submissions in Table~\ref{tab:phase1_final_leaderboards} lie in a narrow score range and are difficult to separate because of the limited number of unseen cosmological models in the holdout set (B). When averaging the scores across (A) and (B), leaderboard (C) shows that \texttt{Shubhojit}, \texttt{Transatlantic Dream Team}, and \texttt{THUML} occupy the top three positions, indicating the strongest balanced performance across both previously seen and unseen cosmological models. 

The noticeable differences in rankings and scores between leaderboards (A) and (B) in Table~\ref{tab:phase1_final_leaderboards} stem from the limited cosmological models in our public datasets, as described in Sec .~\ref{sec:limitations}. As a result, overfitting is more likely to occur in the $101$ cosmological models of leaderboard (A). Nevertheless, all solutions demonstrate strong performance compared to the baseline methods we provided in the competition (see Appendix~\ref{app:baselines} for the details and scores of our baseline methods) in all leaderboards; furthermore, all solutions generalize very well to unseen realizations of systematics, and their corresponding coverages (the chance that the uncertainty estimates cover the ground truths) are generally close to $1\sigma$. 
More details about the final submitted solutions from our participants are available on our website~\cite{Phase1_workshop_agenda}. 
We also summarize the final submissions we received for the final evaluation with their methodologies in Appendix~\ref{app:methods}. Below we present the best solutions.

\begin{table*}[h!]
\centering
\caption{Final leaderboards evaluated on the holdout sets. Holdout set (A) contains new realizations of cosmological models seen in the training and public test data; holdout set (B) contains new realizations of unseen cosmological models. The final aggregate leaderboard uses the average of the scores on (A) and (B).}
\begin{tabular}{llcc}
\hline
Leaderboard & Participant & Score & Standardized MSE / coverage \\
\hline
(A) 1st& \texttt{Cmbagent} & $11.703$ & $0.103$ / $0.700$ \\
(A) 2nd & \texttt{Transatlantic Dream Team} & $11.654$ & $0.104$ / $0.709$ \\
(A) 3rd & \texttt{Shubhojit} & $11.599$ & $0.103$ / $0.658$ \\
\hline
(B) 1st & \texttt{Shubhojit} & $11.361$ & $0.097$ / $0.662$ \\
(B) 2nd & \texttt{THUML} & $11.051$ & $0.105$ / $0.673$ \\
(B) 2nd & \texttt{jagoncalves} & $11.037$ & $0.107$ / $0.668$ \\
(B) 2nd & \texttt{andry834} & $11.001$ & $0.108$ / $0.723$ \\
(B) 2nd & \texttt{jhu\_suicee} & $10.989$ & $0.107$ / $0.645$ \\
(B) 2nd & \texttt{Transatlantic Dream Team} & $10.988$ & $0.107$ / $0.682$ \\
\hline
(C) 1st & \texttt{Shubhojit} & $11.480$ & $0.100$ / $0.660$ \\
(C) 2nd & \texttt{Transatlantic Dream Team} & $11.321$ & $0.106$ / $0.695$ \\
(C) 3rd & \texttt{THUML} & $11.285$ & $0.106$ / $0.679$ \\
\hline
\end{tabular}
\label{tab:phase1_final_leaderboards}
\end{table*}

\subsection{\texttt{Transatlantic Dream Team}: Neural Summaries from Emulated Simulations (NSES)}
\label{subsec:Transatlantic}
We use a Convolutional Neural Network (CNN) to predict the mean and (log-) standard deviation of the cosmological parameters directly from the high-dimensional convergence maps, optimizing the challenge evaluation score as the loss function. As explained earlier, the training dataset contains only $101$ distinct cosmological models and $256$ sets of nuisance parameters per cosmology, making the model highly prone to overfitting. 

To mitigate the data limitation, we combine machine learning (ML) strategies with domain-specific simulation expertise. We explain both aspects below.

\subsubsection{ML Strategies}
On the ML side, we adopt a small EfficientNetV2 architecture \citep{tan2020efficientnetrethinkingmodelscaling, tan_2021} (24 million parameters) and apply extensive data augmentation combined with a careful learning of model parameters.

\paragraph{Data Augmentation} To prevent overfitting, we apply the following augmentations during training: random flips, random cyclic left shifts, and random cyclic upwards shifts. Additionally, we add noise on-the-fly during training to the noise-free convergence maps. This approach enables us to generate as many independent noise realizations as possible to enhance model generalization.

\paragraph{Model Parameters} Model fitting is performed with: (1) the AdamW optimizer \citep{adamw}, (2) dropout \citep{dropout},  (3) weight averaging, and (4) a decaying learning rate across epochs. We motivate the use of the AdamW optimizer and dropout for their regularization effects to prevent overfitting. Two weight averaging schemes were tested: Stochastic Weight Averaging  \citep[SWA,][]{izmailov2019averagingweightsleadswider} and Exponential Moving Average \citep[EMA,][]{tarvainen_2018}. Experimenting with these strategies stabilized the training and validation losses across epochs. Our best score was obtained with EMA. Finally, the learning rate is linearly increased during a warm-up phase and it is then decayed every epoch during the training phase.

\subsubsection{Building Additional Simulations}
We pre-train our EfficientNet on additional simulations that closely resemble the challenge dataset. We first generate a set of LogNormal maps matching the challenge resolution and redshift distribution, then emulate the challenge simulations by learning a pixel-level correction from the LogNormal maps using optimal transport flow-matching.

\begin{figure*}[t!]
\centering
\begin{subfigure}[t]{0.37\textwidth}
    \centering
    \includegraphics[width=\textwidth]{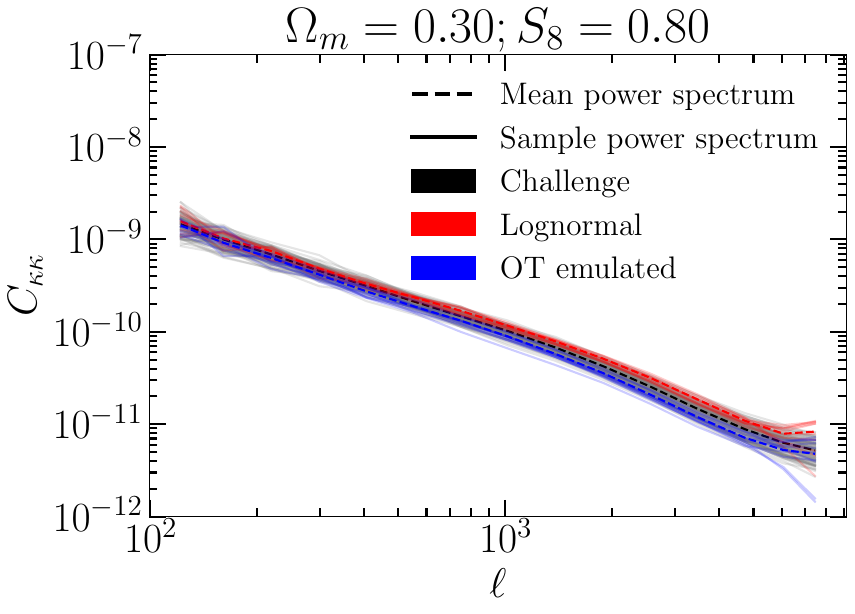}
    \caption{Convergence power spectra from the challenge, LogNormal, and OT-emulated datasets at the same cosmology.}
    \label{fig:ps_comparison}
\end{subfigure}
\hfill
\begin{subfigure}[t]{0.61\textwidth}
    \centering
    \includegraphics[width=\textwidth]{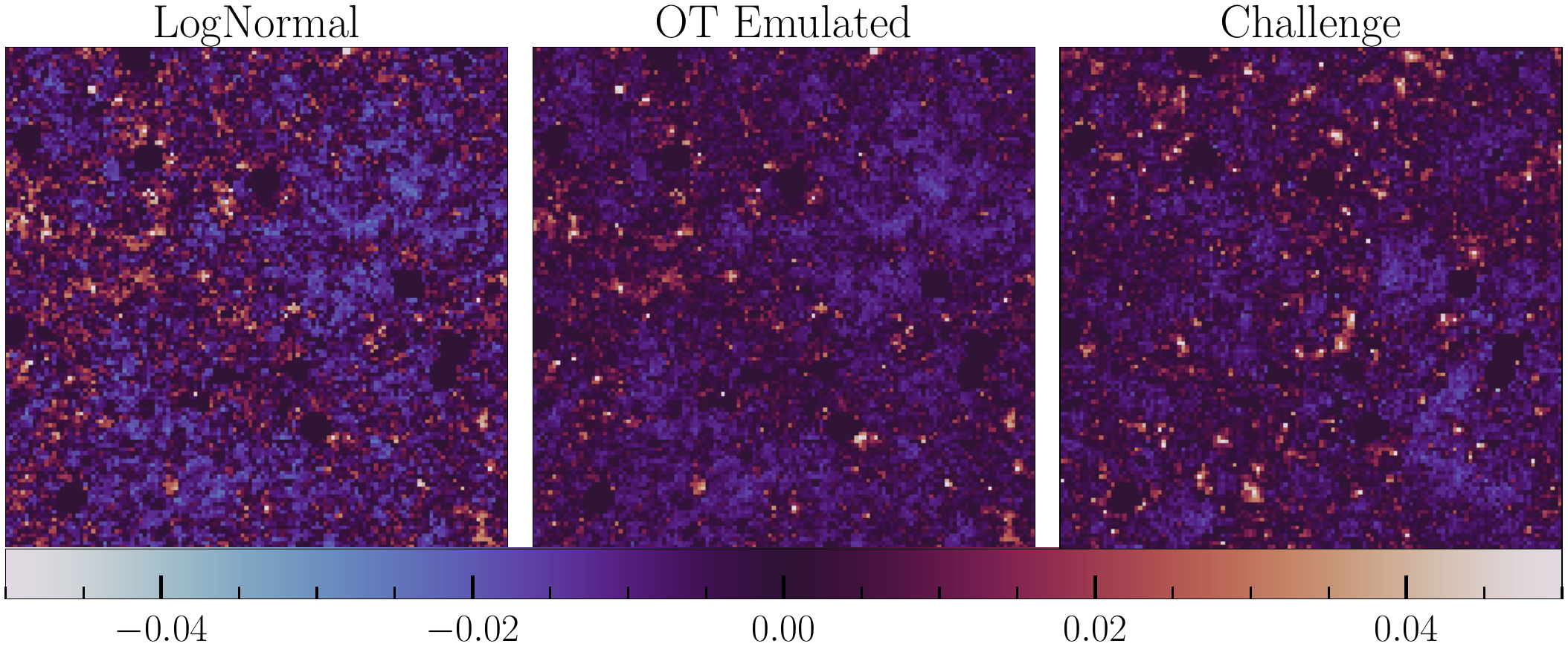}
    \caption{Convergence map comparison: LogNormal (left), OT-emulated (middle), and challenge (right). A $128\times 128$ patch is shown.}
    \label{fig:map_comparison}
\end{subfigure}
\caption{Visual comparison of the simulation datasets at the same cosmology.}
\label{fig:combined}
\end{figure*}

\paragraph{LogNormal Simulations}
We generate cheap LogNormal simulations using the python package \texttt{GLASS} \citep{Tessore_2023}. Convergence HEALPix maps with a resolution of $2$ arcmin are created over cosmological parameters $(\Omega_m, S_8)$. To mimic the challenge dataset, we (i) fix the redshift galaxy distribution to match HSC-Y3's redshift bin 2 and (ii) include the effect of redshift uncertainty in each simulation. A Gnomonic projection is applied to multiple patches of the HEALPix map to obtain 2D convergence images at the challenge resolution. With these settings, we produce 100\,000 simulations over 10\,000 cosmologies in 5 hours on 200 CPUs. Priors are $\Omega_m \sim \mathcal{U}(0.07, 0.65)$, $S_8 \sim \mathcal{U}(0.68, 0.95)$, and $\Delta z \sim \mathcal{N}(0, 0.022^2)$. The cosmological priors are purposefully slightly broader than the challenge's to ease neural network convergence during fine-tuning. We compare the power spectra of convergences from the challenge dataset and from a LogNormal simulation in Fig.~\ref{fig:ps_comparison}; an example LogNormal map is shown in Fig.~\ref{fig:map_comparison}.

\paragraph{Pixel-Level Emulation}

Although LogNormal pre-training improves our score, a mismatch remains: the validation score stays below $-20$, compared to our best score of $11.66$ on challenge simulations. To bridge this gap, we leverage flow-matching conditional optimal transport (COT-FM) \citep{kerrigan2024dynamicconditionaloptimaltransport}, which maps one distribution to another while preserving conditional structure. In particular, \cite{zeghal2025bridgingsimulatorsconditionaloptimal} introduced a pixel-level emulator that uses COT-FM to correct inexpensive simulations towards more accurate ones by transporting simulator likelihoods, applying minimal correction while respecting the conditioning on cosmological parameters. This method operates on unpaired datasets, making it well-suited for this challenge. Due to limited time (COT-FM requires $\approx 17$ hours to train on an NVIDIA A40 GPU), the emulator was not trained long enough to fully map the LogNormal simulations to the data distribution, resulting in a slightly biased power spectrum (Fig.~\ref{fig:ps_comparison}). To pretrain the EfficientNet, we generate $100\,000$ emulated convergence maps. Solving the ordinary differential equation for each LogNormal simulation requires $12$ hours on a NVIDIA A100-SXM4-40GB GPU. This pre-training, combined with a fine-tuning on the challenge data, enabled us to reach a score of $11.6535$ (second place) on (A), $10.9883$ (6th place) on (B), and $11.3209$ (second place) on (C).

\subsection{\texttt{Shubhojit}: Cosmological Parameter Estimation via Parameter-Efficient DenseNet and Tunable Loss Function}

\textbf{Methods:} To efficiently handle the dataset, all 25,856 maps were pre-processed into a single \texttt{float16} memory-mapped file for rapid out-of-core loading. Training utilized on-the-fly augmentations: geometric random flips and physics-based Gaussian noise injection ($n_g=30.0$, pixel size $= 2.0$ arcmin) to faithfully mimic test set noise properties. 

To stabilize training and optimally balance precision with coverage, the standard Negative Log-Likelihood (NLL) was decoupled into a novel, tunable three-term loss function. For a sample $i$, the loss is:
\begin{equation}
    L_i = \lambda_{\log\sigma^2} \underbrace{\sum_{j} \log(\sigma_{ij}^2)}_{\text{Uncertainty Penalty}} + \lambda_{\frac{\text{MSE}}{\sigma^2}} \underbrace{\sum_{j} \frac{(y_{ij} - \mu_{ij})^2}{\sigma_{ij}^2}}_{\text{Weighted MSE}} + \lambda_{\text{MSE}} \underbrace{\sum_{j} (y_{ij} - \mu_{ij})^2}_{\text{Pure MSE}}
\end{equation}
The \textit{Uncertainty Penalty} discourages over-confidence, the \textit{Weighted MSE} penalizes errors when confident, and the \textit{Pure MSE} provides an unweighted gradient anchor. Cross-validation fine-tuned the weights (e.g., $\lambda_{\log\sigma^2}=2.0, \lambda_{\frac{\text{MSE}}{\sigma^2}}=1.0, \lambda_{\text{MSE}}=0.0$). Furthermore, guided initialization set the log-variance output bias to $-5.0$, preventing the variance term from dominating early epochs.

\begin{figure}[t]
    \centering
    \begin{subfigure}[b]{0.48\textwidth}
        \centering
        \begin{tikzpicture}[
            scale=0.65, transform shape, 
            node distance=0.3cm, 
            block/.style={rectangle, draw, fill=blue!20, text width=10em, text centered, minimum height=2em, rounded corners, inner sep=3pt},
            res_block/.style={rectangle, draw, fill=green!30, text width=10em, text centered, minimum height=2em, rounded corners, inner sep=3pt},
            pool_block/.style={rectangle, draw, fill=orange!40, text width=10em, text centered, minimum height=2em, rounded corners, inner sep=3pt},
            head_block/.style={rectangle, draw, fill=purple!30, text width=10em, text centered, minimum height=2em, rounded corners, inner sep=3pt},
            arrow/.style={-Latex, thick},
            data_input/.style={trapezium, trapezium left angle=70, trapezium right angle=110, draw, fill=green!20, text centered, minimum height=2em, inner sep=3pt},
            data_output/.style={trapezium, trapezium left angle=70, trapezium right angle=110, draw, fill=red!30, text centered, minimum height=2em, inner sep=3pt}
        ]
            \node[data_input, text width=9em] (input) {Input Map \\ 1 x 1424 x 176};
            \node[block, below=of input] (conv1) {Initial Conv \\ (Conv, BN, ReLU, Pool)};
            \node[res_block, below=of conv1] (layer1) {ResNet Layer 1 \\ (2x Res Blocks)};
            \node[res_block, below=of layer1] (layer2) {ResNet Layer 2 \\ (2x Res Blocks)};
            \node[res_block, below=of layer2] (layer3) {ResNet Layer 3 \\ (2x Res Blocks)};
            \node[res_block, below=of layer3] (layer4) {ResNet Layer 4 \\ (2x Res Blocks)};
            \node[pool_block, below=of layer4] (pool) {AdaptiveAvgPool2d};
            \node[head_block, below=of pool] (head) {Unified Regressor Head};
            \node[data_output, text width=12em, below=of head] (output) {Final Output \\ $[\mu_{\Omega_m}, \mu_{S_8}, \log\sigma^2_{\Omega_m}, \log\sigma^2_{S_8}]$};

            \draw[arrow] (input) -- (conv1);
            \draw[arrow] (conv1) -- (layer1);
            \draw[arrow] (layer1) -- (layer2);
            \draw[arrow] (layer2) -- (layer3);
            \draw[arrow] (layer3) -- (layer4);
            \draw[arrow] (layer4) -- (pool);
            \draw[arrow] (pool) -- (head);
            \draw[arrow] (head) -- (output);
        \end{tikzpicture}
        \caption{ResNet-based architecture.}
    \end{subfigure}
    \hfill
    \begin{subfigure}[b]{0.48\textwidth}
        \centering
        \begin{tikzpicture}[
            scale=0.65, transform shape, 
            node distance=0.3cm, 
            block/.style={rectangle, draw, fill=blue!20, text width=10em, text centered, minimum height=2em, rounded corners, inner sep=3pt},
            dense_block/.style={rectangle, draw, fill=teal!30, text width=10em, text centered, minimum height=2em, rounded corners, inner sep=3pt},
            trans_block/.style={rectangle, draw, fill=yellow!50, text width=10em, text centered, minimum height=2em, rounded corners, inner sep=3pt},
            pool_block/.style={rectangle, draw, fill=orange!40, text width=10em, text centered, minimum height=2em, rounded corners, inner sep=3pt},
            head_block/.style={rectangle, draw, fill=purple!30, text width=10em, text centered, minimum height=2em, rounded corners, inner sep=3pt},
            arrow/.style={-Latex, thick},
            data_input/.style={trapezium, trapezium left angle=70, trapezium right angle=110, draw, fill=green!20, text centered, minimum height=2em, inner sep=3pt},
            data_output/.style={trapezium, trapezium left angle=70, trapezium right angle=110, draw, fill=red!30, text centered, minimum height=2em, inner sep=3pt}
        ]
            \node[data_input, text width=9em] (input) {Input Map \\ 1 x 1424 x 176};
            \node[block, below=of input] (conv1) {Initial Conv \\ (Conv, BN, ReLU, Pool)};
            \node[dense_block, below=of conv1] (dense1) {Dense Block 1};
            \node[trans_block, below=of dense1] (trans1) {Transition Layer 1};
            \node[dense_block, below=of trans1] (dense2) {Dense Block 2};
            \node[trans_block, below=of dense2] (trans2) {Transition Layer 2};
            \node[dense_block, below=of trans2, text width=10em] (dense_etc) {Dense Blocks 3 \& 4 \\ with Trans Layer 3};
            \node[pool_block, below=of dense_etc] (pool) {AdaptiveAvgPool2d};
            \node[head_block, below=of pool] (head) {Unified Regressor Head};
            \node[data_output, text width=12em, below=of head] (output) {Final Output \\ $[\mu_{\Omega_m}, \mu_{S_8}, \log\sigma^2_{\Omega_m}, \log\sigma^2_{S_8}]$};

            \draw[arrow] (input) -- (conv1);
            \draw[arrow] (conv1) -- (dense1);
            \draw[arrow] (dense1) -- (trans1);
            \draw[arrow] (trans1) -- (dense2);
            \draw[arrow] (dense2) -- (trans2);
            \draw[arrow] (trans2) -- (dense_etc);
            \draw[arrow] (dense_etc) -- (pool);
            \draw[arrow] (pool) -- (head);
            \draw[arrow] (head) -- (output);
        \end{tikzpicture}
        \caption{DenseNet-based architecture.}
    \end{subfigure}
    \vspace{-0.2cm} 
    \caption{The two model architectures are shown in the figure. The models exploit either residual feature skipping or dense feature concatenation to resolve deep maps.}
    \label{fig:sn_architectures}
\end{figure}
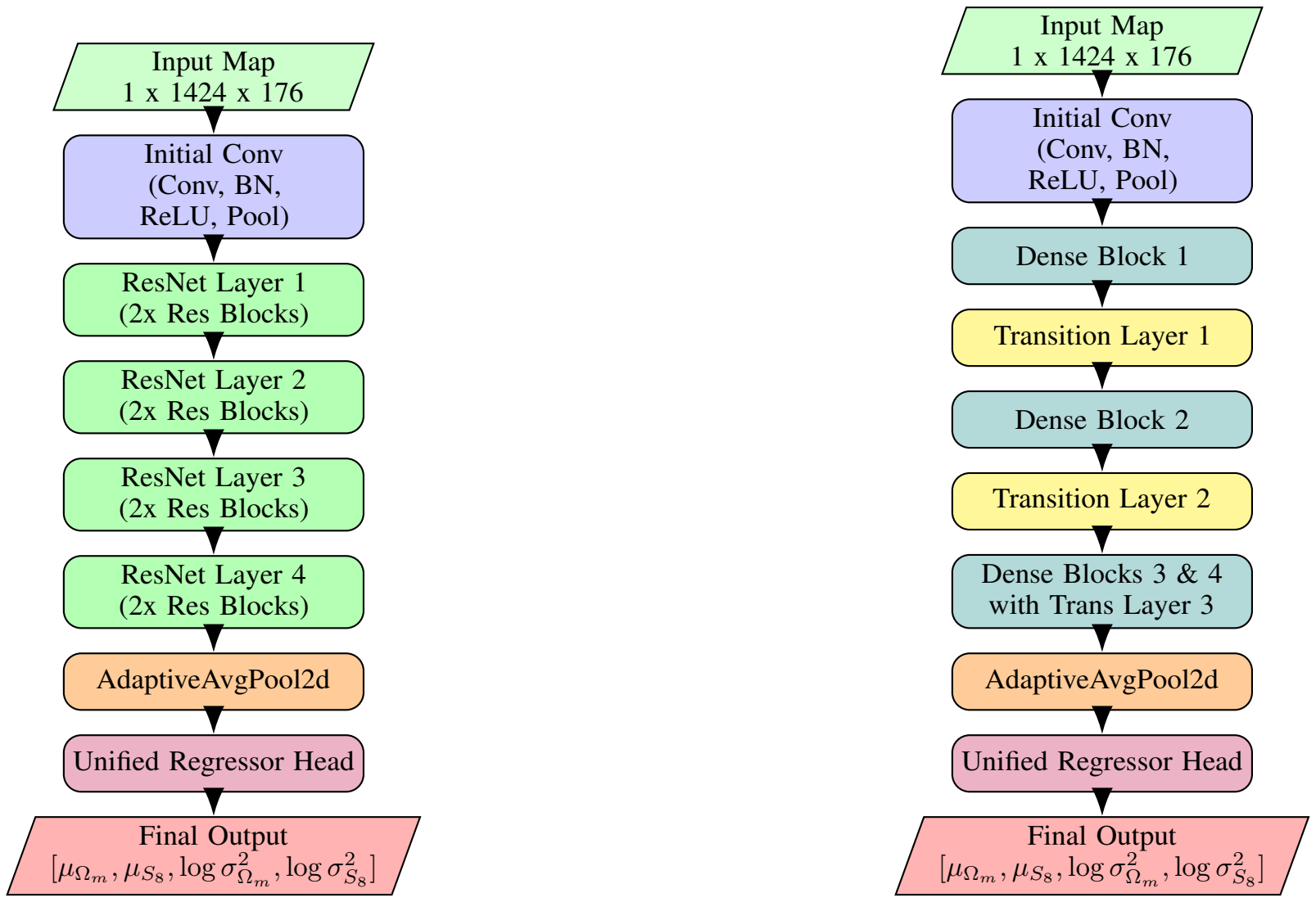

\textbf{Architectural Overview:} Instead of complex multi-head networks, two compact primary architectures were deployed in a 5-fold ensemble (Fig.~\ref{fig:sn_architectures}). Both networks share a macro-topology but utilize different feature extractors: a \textbf{ResNet \cite{he2016deep} backbone} ($\sim$766k parameters, using residual skip connections) and a \textbf{DenseNet \cite{huang2017densely} backbone} ($\sim$472k parameters, using dense blocks for heavy feature concatenation). Both networks feed into an \texttt{AdaptiveAvgPool2d} regularizer followed by a unified regressor head.

\textbf{Results:} The solution evolved systematically from a statistical MCMC emulator baseline (score of $4.54$). Transitioning to a 5-fold CNN ensemble immediately doubled this baseline by leveraging full 2D spatial features. Subsequent architectural enhancements pushed the score to $9.82$. Ultimately, implementing the tunable decoupled loss unlocked the final peak performance, achieving leaderboard scores of $11.48$ for the ResNet ensemble and $11.61$ for the DenseNet ensemble.

\textbf{Key Findings \& Conclusions:} 
\begin{itemize}
    \setlength\itemsep{0em} 
    \item \textit{Loss Decoupling $>$ Architectural Complexity:} Providing granular control over the trade-off between mean prediction accuracy and uncertainty calibration via the decoupled loss yielded significantly greater performance stability than scaling up architectural complexity.
    \item \textit{Parameter Efficiency Generalizes Better:} The highly compact DenseNet model ($\sim$472k params) unexpectedly demonstrated slightly better generalization on the test set compared to the deeper ResNet, suggesting it possesses a more robust inductive bias for evaluating cosmological maps and so DenseNet is chosen for final submission.
    \item \textit{Global Pooling as Regularization:} Using \texttt{AdaptiveAvgPool2d} acts as an effective information bottleneck. It compresses spatial data into salient vectors, implicitly regularizing the network against overfitting.
\end{itemize}

\subsection{\texttt{THUML}: Domain Feature Enhanced Unisolver for Weak Lensing Inference}
\label{subsec:THUML_method}

Drawing inspiration from Unisolver\cite{zhou2024unisolver}, a universal neural PDE solver, we propose a Vision Transformer (ViT) architecture conditioned on handcrafted domain features from cosmology through Adaptive Layer Normalization (AdaLN), and an ensemble selected via greedy search over heterogeneous feature sets and network variants to achieve accurate and robust inference of ($\Omega_m$, $S_8$) from weak lensing convergence maps. Specifically, our method consists of the following three components:

\paragraph{Handcrafted Domain Feature Extraction} Given the rich statistical structure of weak lensing convergence maps, the community has developed a mature set of handcrafted statistical descriptors over the past decades. We utilize domain knowledge and compute a set of classical cosmology-inspired statistical descriptors, including power spectrum, peak counts, Minkowski functionals, etc. These handcrafted features offer interpretable and physically grounded descriptors of both Gaussian and non-Gaussian structures. They serve as stable summary features that complement deep models.

\paragraph{Model Architecture: ViT with AdaLN Conditioning} We design a ViT-based model that processes the full weak lensing convergence map. The key novelty is the integration of physically-motivated, handcrafted features into the deep network via AdaLN, following Unisolver. 

In each Transformer block, given a hidden representation $h$, AdaLN computes:
\[
\mathrm{AdaLN}(h \mid f) = \gamma(f) \cdot \frac{h - \mu(h)}{\sigma(h)} + \beta(f)
\]
where $f$ denotes the vector of handcrafted features, and $\gamma(f), \beta(f)$ are learned affine transformations of $f$. This modulation mechanism allows the network to adaptively rescale and shift its internal representations based on the global physical context encoded in $f$.

Concretely, the convergence map is first partitioned into patches and embedded into a latent space with positional encoding, while the handcrafted features are projected into a conditioning embedding that is injected throughout the network via AdaLN. A stack of Transformer blocks with self-attention and feed-forward layers, combined with residual connections and gating, processes these representations. And a final conditioned projection outputs the target cosmological parameters along with their uncertainties. This design allows the model to jointly exploit global physical descriptors and high-dimensional learned features, improving both generalization and interpretability.

\paragraph{Model Ensemble} We build an ensemble by training multiple model variants that differ along two dimensions: combinations of handcrafted features and model architectures. Specifically, we consider various feature sets composed of power spectrum, peak counts, Minkowski functionals, and other domain-specific descriptors. In addition to the standard architecture, we explore two variants within the same framework: one that incorporates a lightweight CNN front-end to extract low-level representations from convergence maps, and another that operates solely on handcrafted features without taking convergence maps as input.

Each candidate model is evaluated on the validation set. We then perform a greedy ensemble search that iteratively selects the model which yields the largest improvement in the validation score when combined with the current ensemble. The ensemble result is obtained by taking the average of the predictions from all candidate models. This process continues until no further gain can be achieved. The resulting ensemble typically consists of complementary models that cover diverse feature combinations and architectural choices. Based on the procedures described above, we find a highly effective subset that includes 16 models to form our final ensemble.

\paragraph{Results} Using the ensemble constructed by the above procedure, we achieve a score of $11.5185$ on the in-distribution test set and $11.0511$ on the out-of-distribution test set (unseen cosmological parameters). These results highlight the effectiveness of integrating domain knowledge, deep architectures, and ensemble modeling for robust and uncertainty-aware inference in weak lensing cosmology.

\subsection{\texttt{Cmbagent}: Ensembling Inception CNNs and Dihedral Augmentation}
\label{subsec:cmbagent}
For details on this method we refer the reader to \textit{Borrett et al} \cite{borrett2026}. The particularity of this method is that it was agent-driven, relying primarily on the \href{https://github.com/CMBAgents/cmbagent}{\texttt{Cmbagent}}\footnote{https://github.com/CMBAgents/cmbagent} \textit{Planning \& Control} framework \cite{Laverick:2024fyh, CMBAGENT_2025, xu2025, villaescusanavarro2025denarioprojectdeepknowledge}. The team leveraged the power of Large Language Models (LLM) to generate ideas on how to solve the problem and implement them. This method allowed the team to reach the top of the public leaderboard (first 10 entries out of 1759) and the first-place on leaderboard (A). The list of concepts or solutions that were suggested and implemented by agents and enabled a significant advantage are:
\begin{itemize}
    \item Inception-style CNN architecture \citep{Szegedy2015}, including Inception Squeeze-and-Excitation \citep{hu2019squeezeandexcitationnetworks}. 
    \item Dihedral group of transformations. 
    \item Ensemble Weighting by Negative Log-Likelihood.
    \item Grid-based inference with a Ledoit--Wolf-style shrunk covariance estimator \citep{LedoitWolf2003}. 
    \item Bias-correcting loss for the final hybrid network.
\end{itemize}

The model belongs to the \texttt{Inception} family of CNN architectures \citep{Szegedy2015} with \textit{Squeeze and Excitation} (SE) blocks referred to as \texttt{InceptionSE} \citep{hu2019squeezeandexcitationnetworks}. More precisely, the model is composed of an ensemble of 10 CNNs (5  \texttt{InceptionSE} and 5  \texttt{Inception}), i.e., a 10-ensemble. 

The 10-ensemble CNN was trained on the entire Challenge dataset, with an 80/20 training-validation split with an MSE loss, for 15 epochs on noiseless maps followed by 50 epochs on noisy maps.  The data was massively augmented, first with noise resampling at each epoch and on-the-fly, then second with all transformations of $D_4$ (the dihedral group).

On top of the 10-ensemble CNNs described above, we added scattering covariances \cite{2012arXiv1203.1513B,Allys2020,Cheng2024}. This was done in two stages. First, we reduced the dimensionality of the ST features to 64 using principal component analysis (PCA) and trained a lightweight regression multilayer perceptron (MLP), adding its weighted output to the CNN prediction. We additionally added an $L_2$ penalty on the deviation from the CNN baseline and an auxiliary classification term and MLP to predict the error direction. Finally, we fine-tuned the combined ensemble for 25 epochs with a reduced learning rate to maintain stability while leveraging the ST statistics.  All SC coefficients were computed using the $\texttt{foscat}$\footnote{https://github.com/jmdelouis/FOSCAT} package, which supports convolutions on rectangular and masked maps.

We trained the model on a single Nvidia RTX 6000 Pro Blackwell. Training a single \texttt{Inception} or \texttt{InceptionSE} ensemble takes approximately 30 minutes, for a total of about 5 hours to train the 10 ensembles ultimately used for inference. The inference stage requires roughly 8 minutes when using all 10 ensembles for 4000 test maps. Our \texttt{Inception} model contains 422,754 parameters, while \texttt{InceptionSE} contains 722,610 parameters. Using single-precision floating point format (FP32 precision), these correspond to storage footprints of 1.7 MB and 2.9 MB per model. Introducing the ST MLP heads raises this by a small amount to 431,784 parameters for the \texttt{Inception} model and 731,640 for \texttt{InceptionSE}. The method utilizes a total dataset of 830 GB, dominated by pre-generated noisy realizations of weak lensing convergence maps (5.5 GB per realization).

Our inference pipeline also differs from the direct prediction and Markov Chain Monte Carlo (MCMC) scripts provided in the challenge starting kit. The approach we utilize, suggested by \texttt{GPT-5} \cite{singh2025openaigpt5card}, aggregates validation predictions from multiple ensemble members, smooths them across a cosmology grid, and uses them to define a Gaussian likelihood model. A ten-fold cross-validation hyperparameter survey was carried out by \href{https://github.com/CMBAgents/cmbagent}{\texttt{Cmbagent}} to optimize this procedure. For inference on the test data, ensemble-weighted model outputs are passed through the calibrated likelihood to obtain final posterior means and uncertainties for each map. 

\begin{table}[t]
\centering
\caption{Summary of model–generation pathways used during the architecture search, showing the models proposed under different \href{https://github.com/CMBAgents/cmbagent}{\texttt{Cmbagent}}/LLM configurations and their corresponding test-set performance. The left column indicates the source of each proposed architecture: some models were generated directly by \href{https://github.com/CMBAgents/cmbagent}{\texttt{Cmbagent}} (either through \textit{Planning \& Control} (P\&C) or \textit{One-shot}), while in other cases a base model first proposed by \href{https://github.com/CMBAgents/cmbagent}{\texttt{Cmbagent}} was subsequently refined by external LLMs such as \texttt{GPT-5} or \texttt{Gemini} \cite{comanici2025gemini25pushingfrontier}. In the inference method column, “MCMC’’ refers to sampling the posterior using a Markov Chain Monte Carlo algorithm. “Grid’’ means evaluating the likelihood on a fixed parameter grid and forming parameter estimates by weighting grid points according to their likelihood values. \texttt{BaseCNN} denotes our modified version of the starter-kit \texttt{SimpleCNN}, with slightly wider convolutional layers and a replacement of the large fully connected head by global average pooling followed by a small MLP head.
}
\begin{tabular}{l c c c c}
\hline
\textbf{Agents/LLMs} & \textbf{Model} & \textbf{Inference} &
\textbf{Augmentation} & \textbf{Score} \\
\hline
\texttt{Cmbagent}-P\&C & ResNet-18/34/50 & MCMC & Flips/rotations & $\approx 10.5$ \\
+ \texttt{Cmbagent}-One shot & BaseCNN & MCMC & No & $11.02$ \\
+ \texttt{GPT-5} & BaseCNN & Grid & No & $11.42$ \\
+ \texttt{Gemini-2.5-Pro} & Inception + InceptionSE & Grid & $D_4$ & $11.72$ \\
+ Humans & WST & Grid & $D_4$ & $11.73$ \\
\hline
\end{tabular}
\label{tbl:llm_results}
\end{table}

Table \ref{tbl:llm_results} summarizes the sequence of modeling strategies we evaluated, ranging from purely \href{https://github.com/CMBAgents/cmbagent}{\texttt{Cmbagent}}-generated architectures (from either \textit{Planning \& Control} or \textit{One-shot} mode) to models refined by additional LLMs and human modification. The table highlights how progressively adding these refinement stages affects the resulting predictive performance. Note that after each proposed method, we updated the prompts provided to \href{https://github.com/CMBAgents/cmbagent}{\texttt{Cmbagent}} and the external LLMs so that each subsequent model was generated with awareness of the previous design. 

Although the final model performed remarkably well on the public leaderboard and leaderboard (A), it is worth noting that it did not rank in the top three places of leaderboard (B). The team had not anticipated that the public leaderboard contained the same cosmological models than those of the training dataset provided to the participants. Furthermore, they followed the recommendation of the starting kit encouraging a training-validation split along the nuisance parameters axis. This led to an overfitting of the original cosmology and poor results on leaderboard (B).

\section{Conclusions and Outlook}

We present the FAIR Universe Weak Lensing ML Uncertainty Challenge, the first benchmark and ML competition for cosmological inference from weak gravitational lensing convergence maps under realistic systematics. The dataset combines high-resolution N-body simulations and ray-tracing with systematic bias from baryonic feedback and photometric redshift uncertainty, to closely emulate the conditions of real survey analyses.

A central limitation of the benchmark is the modest number of cosmological models, which is dictated by the cost of the underlying simulations. We argue this constraint reflects rather than diminishes the benchmark's value: realistic simulation-based inference in cosmology is itself rate-limited by simulation cost, and a key goal of our competition is to identify methods that extract information most efficiently from limited training data. 

Building on lessons from Phase 1 competition, we have expanded the simulation suite with new cosmological models. The new simulations are used to construct a larger test set drawn entirely from cosmological models not in the training data. To preserve the integrity of the ongoing Phase 2 competition, this test set will be released publicly only after Phase 2 concludes.

We evaluate the performance of novel ML methods developed by the participants of our competition. The results show that several ingredients are important in reaching strong cosmological inference performance, such as domain-specific data augmentation, ensembles of compact and well-regularized neural network architectures, and the integration of domain knowledge through handcrafted features. The winning solutions further demonstrate that simulator bridging with optimal transport, hybrid inference pipelines, and AI agent-assisted model development workflows can all be effective when paired with careful validation. 
We view this challenge as a community testbed for systematic comparison of cosmological inference methods under controlled but nontrivial conditions. More broadly, we believe this will guide next-generation weak lensing pipelines and improve the reliability of ML inference approaches for future cosmological surveys.


\begin{ack}
We are grateful to the US Department of Energy, Office of High Energy Physics, subprogram on Computational High Energy Physics, for sponsoring this research. 
B.D. acknowledges support from the Ambrose Monell Foundation, the Corning Glass Works Foundation Fellowship Fund, and the Institute for Advanced Study. This research used resources of the National Energy Research Scientific Computing Center (NERSC), a Department of Energy User Facility, using NERSC award HEP-ERCAP0032917.
B.R. gratefully acknowledges the support of the NSF-Simons AI-Institute for the Sky (SkAI) via grants NSF AST-2421845 and Simons Foundation MPS-AI-00010513. This research used the DeltaAI advanced computing and data resource, which is supported by the National Science Foundation (award OAC 2320345) and the State of Illinois. DeltaAI is a joint effort of the University of Illinois Urbana-Champaign and its National Center for Supercomputing Applications
\end{ack}

\clearpage

\bibliographystyle{tepml}
\bibliography{ref.bib,HEPML}
\clearpage


\appendix

\section{Simulation details}
\label{app:simulation}

\begin{table}[b]
\centering
\caption{N-body simulation parameters for different redshift ranges.\label{tab:nbody}}
\begin{tabular}{c|ccccc}
\hline
redshift & box size & $N_{\mathrm{particle}}$ & force resolution & $N_{\mathrm{step}}$ & simulation code\\
\hline
$0<z<0.45$ & $320\ h^{-1}\mathrm{Mpc}$ & $960^3$ & $0.03\ h^{-1}\mathrm{Mpc}$ & adaptive & MP-Gadget \cite{yu_feng_2018_1451799} \\
$0.45<z<1.05$ & $704\ h^{-1}\mathrm{Mpc}$ & $2816^3$ & $0.125\ h^{-1}\mathrm{Mpc}$ & $60$ & FastPM \cite{Feng:2016yqz} \\
$1.05<z<2.72$ & $1536\ h^{-1}\mathrm{Mpc}$ & $1536^3$ & $0.5\ h^{-1}\mathrm{Mpc}$ & $15$ & FastPM \cite{Feng:2016yqz} \\
\hline
\end{tabular}
\end{table}

We construct the dataset using high-resolution cosmological N-body simulations \cite{yu_feng_2018_1451799,Feng:2016yqz} combined with a ray-tracing pipeline. The simulations adopt the survey footprint, source redshift distribution, and shape-noise properties of the HSC Three-Year shape catalogue \cite{Li:2021mvq,rau2023weak}. To capture both large-scale modes at high redshift and small-scale clustering at low redshift within a manageable computational budget, each cosmology is simulated with three nested N-body runs of different box sizes and mass resolutions, each covering a distinct redshift range. The simulation parameters are summarized in Table~\ref{tab:nbody}. Each simulation takes about 20 thousand CPU hours.

A total of 43 snapshots, spaced by $100~h^{-1}\mathrm{Mpc}$ in comoving distance, are saved across the three boxes and used as lens planes. The volume remapping technique \cite{carlson2010embedding} is applied to deform the periodic simulation volume into the elongated geometry of the HSC Y3 footprint, after which a multi-plane ray-tracing algorithm \cite{jain2000ray, hilbert2009ray, Petri2016mocking} computes the weak lensing signal under the flat-sky approximation. To increase the effective number of independent realizations per cosmology, each lens plane is independently translated, rotated, and reflected at every redshift before being stacked along the line of sight \cite{petri2016sample}.

\begin{figure}[t]
    \centering
    \includegraphics[width=0.9\linewidth]{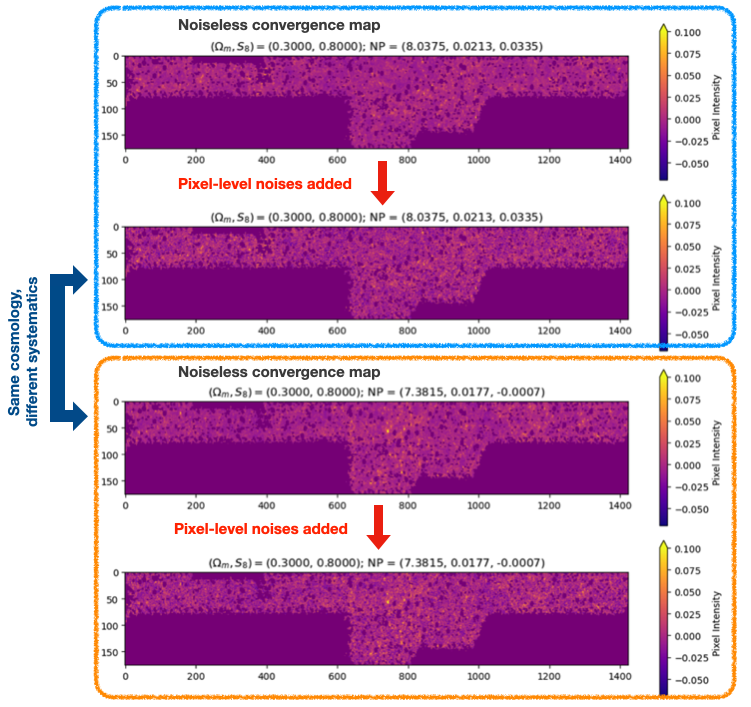}
    \caption{Example noiseless and noisy weak lensing convergence maps. The top panel and bottom panel show two different maps generated with different parameters of systematic effects and random seed, assuming the same cosmological parameters.}
    \label{fig:data_shift}
    \vspace{-0.2cm}
\end{figure}

The HSC Y3 galaxy catalog spans six disjoint sky regions and four tomographic redshift bins. To reduce the size of the released dataset, we restrict the simulation to a single region (WIDE12H) and a single tomographic bin (the second). From the simulated galaxy catalogs, we estimate convergence maps at $2~\mathrm{arcmin}$ resolution, which form the input data for the challenge. Example convergence maps are shown in Fig.~\ref{fig:data_shift}.

\begin{figure}[h!]
    \centering
    \includegraphics[width=0.6\linewidth]{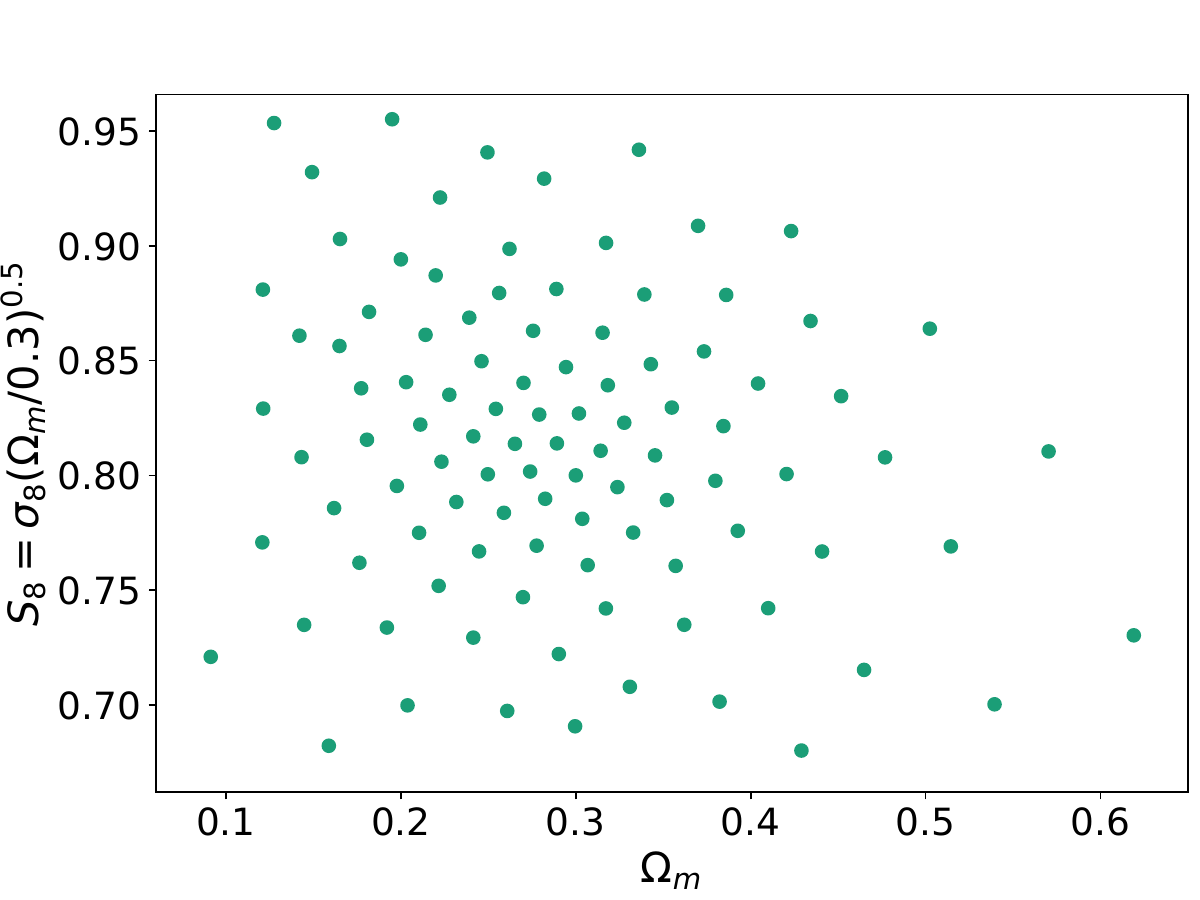}
    \caption{The $101$ cosmological parameters ($\Omega_m, S_8$) in our suite of weak lensing simulations.}
    \label{fig:cosmology}
    \vspace{-0.2cm}
\end{figure}

Each of the $101$ simulations corresponds to a spatially-flat $\Lambda$CDM cosmology that varies in $\Omega_m$ and $\sigma_8$, while the other cosmological parameters are fixed at $\Omega_b = 0.046$, $h = 0.7$, and $n_s = 0.97$. The two parameters of interest, $\Omega_m$ and $S_8 = \sigma_8 (\Omega_m / 0.3)^{0.5}$, are shown in Fig.~\ref{fig:cosmology}. The sampling in this plane is non-uniform, with sample density increasing toward the fiducial values $\Omega_m = 0.3$ and $S_8 = 0.8$ to reduce model interpolation error around currently favored cosmologies. The parameter range is chosen to cover the posterior distributions reported by recent surveys, including HSC Y3 \cite{li2023hyper,dalal2023hyper}, DES Y3 \cite{abbott2022dark}, and KiDS-1000 \cite{asgari2021kids}.

We model the impact of baryonic physics on the matter distribution by applying a transfer function $T(k, z)$ to the simulated density field \cite{sharma2025field}:
\begin{equation}
    \delta(k, z) \;\rightarrow\; \delta(k, z)\, T(k, z),
\end{equation}
where $\delta(k)$ is the Fourier transform of the matter overdensity field $\delta(x) = \rho(x)/\bar{\rho}(x) - 1$, and $\rho(x)$ is the density map from the N-body simulation. The transfer function $T(k, z)$ suppresses small-scale modes and mimics AGN feedback. It is modeled using the halo-model framework of HMcode \cite{mead2021hmcode}, with two free parameters: $T_{\mathrm{AGN}}$, controlling the strength of AGN feedback, and $f_0$, controlling star formation. These nuisance parameters are sampled uniformly in the ranges $T_{\mathrm{AGN}} \in [7.2, 8.5]$ and $f_0 \in [0, 0.0265]$.

The uncertainty in the inferred redshift distribution of source galaxies is modeled as a coherent shift of the redshift distribution:
\begin{equation}
    p(z) \;\rightarrow\; p(z + \Delta_z),
\end{equation}
where $p(z)$ is the redshift distribution of the second tomographic bin of HSC Y3, and $\Delta_z$ is drawn from a Gaussian prior with zero mean and standard deviation $\sigma_{\Delta_z} = 0.022$ \cite{rau2023weak}.

All three nuisance parameters $(T_{\mathrm{AGN}}, f_0, \Delta_z)$ are fully sampled in the training set, allowing participants to marginalize over them during inference.

Measurement noise from the random orientations of source galaxy shapes is modeled as pixel-wise Gaussian noise with zero mean and standard deviation
\begin{equation}
\sigma = \frac{\sigma_\epsilon}{\sqrt{2 n_{\mathrm{g}} A_{\mathrm{pixel}}}},
\end{equation}
where $\sigma_\epsilon \approx 0.4$ is the typical intrinsic ellipticity of a galaxy, $A_{\mathrm{pixel}} = 4~\mathrm{arcmin}^2$ is the pixel area, and $n_g = 30~\mathrm{arcmin}^{-2}$ is the assumed source galaxy number density. This galaxy density is higher than the value reached by current Stage-III weak lensing surveys and is chosen to be representative of the sensitivity expected from upcoming Stage-IV surveys such as Rubin/LSST and Euclid. Since the noise is independent across pixels, the dataset is released without noise added, and shape noise is applied as a lightweight on-the-fly augmentation during training.

The full training set comprises $256$ independent realizations per cosmology, with each realization sampling a fresh draw of the three nuisance parameters and an independent random seed for the lensing maps. Across the $101$ cosmological models, this yields a four-dimensional training tensor of shape $101 \times 256 \times 1424 \times 176$, for a total of $25{,}856$ maps. Every map is annotated with a five-dimensional label that records both the underlying cosmology, $(\Omega_m, S_8)$, and the nuisance configuration used to generate it, $(T_{\mathrm{AGN}}, f_0, \Delta_z)$.

\clearpage

\section{Baseline methods}
\label{app:baselines}
We provide several starting kit notebooks on our \href{https://github.com/FAIR-Universe/Cosmology_Challenge}{GitHub repository}~\github{FAIR-Universe/Cosmology_Challenge}. The starting kits include simple scripts for loading and visualizing data, training and evaluating baseline methods, and preparing a submission file for the competition. The starting kits are designed to enable participants to easily get familiar with the dataset, the traditional methods for the challenge tasks, and the submission workflow. The baseline methods are used to set the baseline scores for the competition.

The baseline methods include binning the power spectrum (a two-point correlation function of matter density fluctuations commonly used in cosmology) as a summary statistic or training a convolutional neural network to compress the data for downstream tasks or direct predictions.

\subsection{Power spectrum analysis with MCMC}
\label{sec:PS_MCMC}

In cosmology, the power spectrum describes how matter is distributed across different size scales in the universe and is a key tool for studying the growth of cosmic structure. Starting from the convergence map $\kappa(\boldsymbol{\theta})$ (a 2D projection of the matter overdensity weighted by lensing geometry), where $\boldsymbol{\theta}$ is the angular coordinate of the observed sky patch, we transform it into Fourier space to get 
\begin{equation}
    \tilde{\kappa}(\boldsymbol{k})
    = \int d^2\boldsymbol{\theta} \, 
      \kappa(\boldsymbol{\theta}) \,
      e^{-i\boldsymbol{k} \cdot \boldsymbol{\theta}}~,
\end{equation}
which represents fluctuations as waves of different wavelengths. The matter power spectrum $P_k$ is then defined by:
\begin{equation}
    \langle \tilde{\kappa}(\mathbf{k}) \tilde{\kappa}^*(\mathbf{k}') \rangle = (2\pi)^2 \delta_D(\mathbf{k}-\mathbf{k}') P_k~,
\end{equation}
where $k = |\mathbf{k}|$ is the wavenumber corresponding to a scale $\lambda \sim 1/k$, and $ \delta_D$ is the Dirac delta function. We can estimate the power spectrum by averaging over all Fourier modes within a wavenumber bin
\begin{equation}
    P_k = 
    \left\langle |\tilde{\kappa}(\boldsymbol{k})|^2 \right\rangle_{k \in \mathrm{bin}}~. \label{eq:Pk}
\end{equation}
Intuitively, $P_k$ tells us how ``clumpy'' the universe is on different scales. In cosmology, the shape and amplitude of $P_k$ encode the physics and composition of the universe, making it one of the most important statistical tools in the field. 

Given a convergence map generated from a simulation with cosmological parameters $\boldsymbol{\Theta} = (\Omega_m, S_8)$, the likelihood of an observation represented by a summary statistic $\boldsymbol{d}_{\rm obs}$ can be modeled as a Gaussian likelihood
\begin{equation}
    p(\boldsymbol{d}_{\rm obs}|\boldsymbol{\Theta}) \propto \frac{1}{\sqrt{|{\rm Cov}(\boldsymbol{\Theta})|}} \exp \left\{-\frac{1}{2}[\boldsymbol{d}_{\rm obs}-\mu(\boldsymbol{\Theta})]^T {\rm Cov}^{-1}(\boldsymbol{\Theta})[\boldsymbol{d}_{\rm obs}-\mu(\boldsymbol{\Theta})]\right\}~.     \label{eq:gauss_lh}
\end{equation}
When there are $N_{\rm sys}$ realizations (parameterized by nuisance parameters $\boldsymbol{\alpha}$) for each cosmological model, the mean summary statistic evaluated at $\boldsymbol{\Theta}$ is
\begin{equation}
    \mu(\boldsymbol{\Theta}) =\frac{1}{N_{\rm sys}} \sum_{j=1}^{N_{\rm sys}} \boldsymbol{d}(\boldsymbol{\Theta},\boldsymbol{\alpha}_j)~, \label{eq:mu}
\end{equation}
where $\boldsymbol{d}(\boldsymbol{\Theta},\boldsymbol{\alpha}_j)$ is the summary statistic of the $j$-th realization of cosmology $\boldsymbol{\Theta}$, and the covariance matrix of the summary statistic is
\begin{equation}
    \operatorname{Cov}(\boldsymbol{\Theta})=\frac{1}{N_{\rm sys}-n_d-2} \sum_{j=1}^{N_{\rm sys}}~[\boldsymbol{d}(\boldsymbol{\Theta},\boldsymbol{\alpha}_j)-\mu(\boldsymbol{\Theta})]^T ~[\boldsymbol{d}(\boldsymbol{\Theta},\boldsymbol{\alpha}_j)-\mu(\boldsymbol{\Theta})]~,    \label{eq:cov}
\end{equation}
where $n_d$ is the number of summary statistics.

In the standard power spectrum analysis, the logarithm of the power spectrum in Eq.~(\ref{eq:Pk}) is treated as a summary statistic for Eq.~(\ref{eq:gauss_lh}) to estimate the cosmological parameters
\begin{equation}
    \boldsymbol{d}(\boldsymbol{\Theta}, \boldsymbol{\alpha}) = \text{log} ~P_k(\boldsymbol{\Theta}, \boldsymbol{\alpha})~,  \label{eq:summary_statistic_PS}
\end{equation}
where $n_d$ is the number of bins in wavenumber $k$. 

The power spectrum emulator $P_k(\boldsymbol{\Theta}, \boldsymbol{\alpha})$ can be obtained using the training dataset we provide. At the inference time, we sample the posterior $p(\boldsymbol{\Theta}|\boldsymbol{d}_{\rm obs})$ with MCMC using the likelihood function (\ref{eq:gauss_lh}) and a flat prior of $\boldsymbol{\Theta}$, and compute the the point estimates $\hat{\boldsymbol{\Theta}} = (\hat{\Omega}_m$, $\hat{S}_8)$ and their uncertainties $(\hat{\sigma}_{\Omega_m}$, $\hat{\sigma}_{S_8})$ from the posterior samples. 

In our starting kit, we also show that the OoD detection probability can be conservatively estimated by comparing the dataset likelihood with the distribution of the training data likelihood.

\begin{figure}[t]
    \centering
    \includegraphics[width=0.8\linewidth]{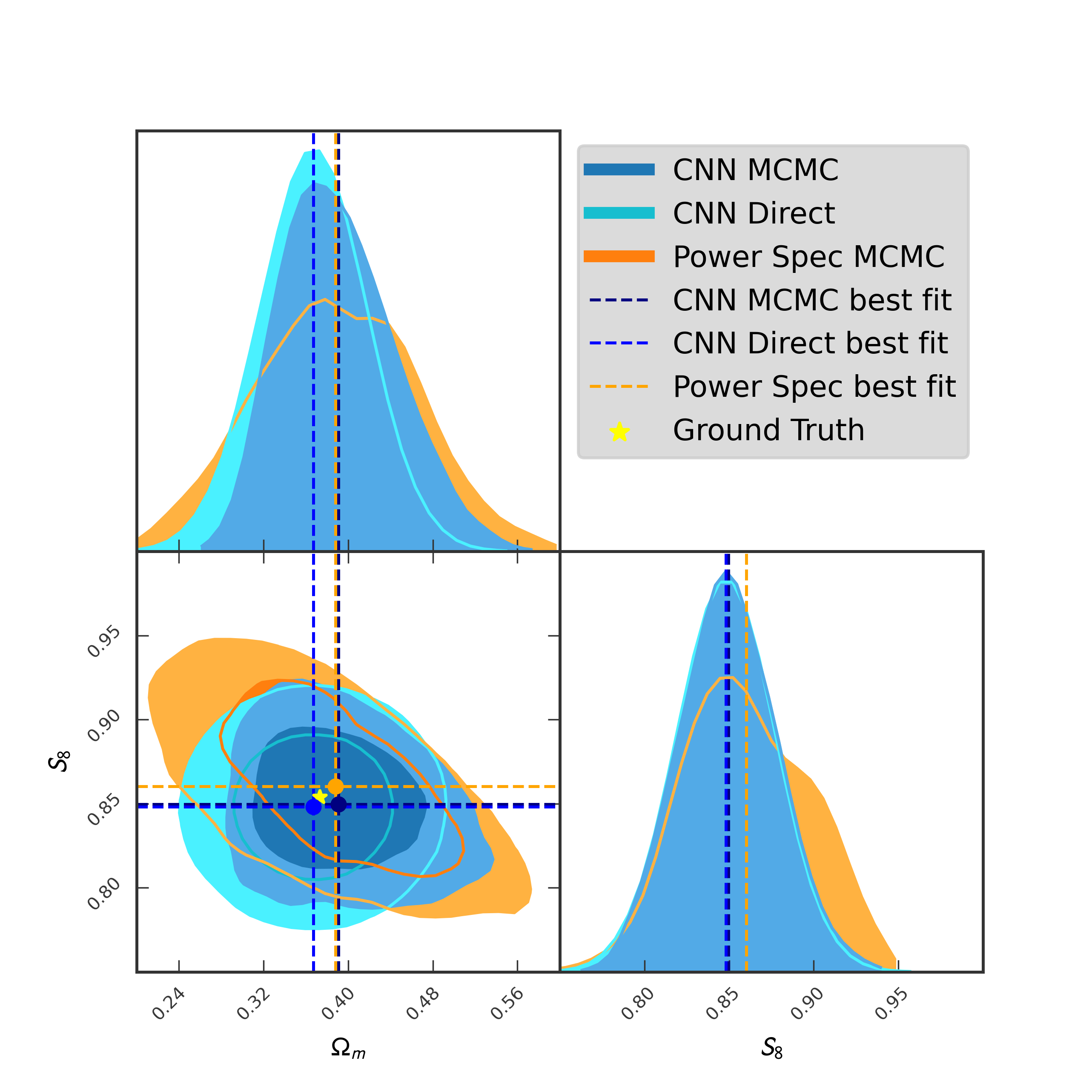}
    \caption{A comparison between the posterior distributions inferred by our baseline methods. \textit{The neural network-based approaches can  estimate the parameters more accurately.}}
    \label{fig:p1_baseline_comp}
    \vspace{-0.2cm}
\end{figure}

\subsection{Convolutional neural network with MCMC}
\label{sec:CNN_MCMC}

Neural networks are very powerful in extracting useful features from high-dimensional data. In this baseline method, we model the mapping from convergence maps to the cosmological parameters $(\Omega_m, S_8)$ using a convolutional neural network (CNN) implemented in \textsc{PyTorch}~\cite{pytorch}.

The input of the CNN is a single–channel 2D convergence map of $1424 \times 176$ pixels. The architecture consists of four convolutional blocks, each composed of a 2D convolution, batch normalization, ReLU activation, and $2\times 2$ max pooling. The numbers of feature channels are $16$, $32$, $64$, and $128$ for the four convolutional blocks, respectively. The first convolution uses a $5\times 5$ kernel with stride $2$ and padding $2$, while the remaining convolutions use $3\times 3$ kernels with stride $1$ and padding $1$.

The output of the convolutional stack is then flattened and passed through a fully connected regression head: two linear layers with $512$ and $128$ units, each followed by ReLU activation and dropout with probabilities $0.2$ and $0.1$, respectively, and a final linear layer outputting the two target parameters.

The CNN is trained using mean–squared error (MSE) loss between the predicted and true $(\Omega_m, S_8)$ values. Optimization is performed using \texttt{Adam} with a learning rate of $2\times 10^{-4}$, weight decay of $10^{-4}$, batch size of $64$, and for a total of $15$ epochs. We apply a \texttt{ReduceLROnPlateau} scheduler that decreases the learning rate by a factor of $0.5$ if the validation loss does not improve for five consecutive epochs. The training is performed using GPUs when available.

The outputs of the trained CNN are the point estimates $\hat{\boldsymbol{\Theta}} = (\hat{\Omega}_m, \hat{S}_8)$, which can be treated as the summary statistic for Eq.~(\ref{eq:gauss_lh}) 
\begin{equation}
    \boldsymbol{d}(\boldsymbol{\Theta}, \boldsymbol{\alpha}) = f_{\rm NN}^{\phi}\left(\kappa(\boldsymbol{\Theta}, \boldsymbol{\alpha})\right) =\hat{\boldsymbol{\Theta}} = (\hat{\Omega}_m, \hat{S}_8)~,  \label{eq:summary_statistic_CNN}
\end{equation}
where $n_d = 2$. The MCMC is then applied to generate samples of the posterior distribution that can be used to compute the uncertainties.

\subsection{Convolutional neural network direct prediction}
\label{sec:CNN_dir}

In this baseline method, we do not fit any predefined summary statistic to the data and perform MCMC. Instead, we estimate the uncertainties directly using the CNN. This is achieved by optimizing a KL divergence objective function using neural network predictions during training
\begin{equation}
    \mathcal{L}_{\rm KL} = \frac{1}{N} \sum_i^{N}\left\{\frac{\left(\hat{\Omega}_{m, i}-\Omega_{m, i}^{\text {truth }}\right)^2}{\hat{\sigma}_{\Omega_m, i}^2}+\frac{\left(\hat{S}_{8, i}-S_{8, i}^{\text {truth }}\right)^2}{\hat{\sigma}_{S_8, i}^2}+\log \left(\hat{\sigma}_{\Omega_m, i}^2\right)+\log \left(\hat{\sigma}_{S_8, i}^2\right)\right\}~.    
\end{equation}

We use the same CNN architecture and hyperparameters described in Sec.~\ref{sec:CNN_MCMC}, except that we change the output dimension of the final layer to $4$ so that for each 2D convergence map the CNN will predict its cosmological parameters $(\hat{\Omega}_m, \hat{S}_8)$ and the standard deviations of the joint Gaussian posterior distribution $(\hat{\sigma}_{\Omega_m}, \hat{\sigma}_{S_8})$. The advantage of this approach is that it directly forces the CNN to learn to predict the point estimates and uncertainties optimizing the scoring function defined in Eq.~(\ref{eq:inf_score}) while avoiding choosing any summary statistic and saving lots of time from MCMC sampling.

\subsection{Baseline scores}
Figure~\ref{fig:p1_baseline_comp} shows a comparison between the sampled posterior distributions of our three baseline methods. Using the scoring metric defined in Eq.~(\ref{eq:inf_score}), the power spectrum analysis, CNN with MCMC, and CNN direct prediction obtain scores of $4.5766$, $8.6809$, and $8.5209$, respectively, on the public test dataset. Compared to the traditional power spectrum analysis, the neural networks can capture more information from the weak lensing data, leading to better predictions. 

\clearpage

\section{Summary of final submissions}
\label{app:methods}

\begin{figure}[h]
    \centering
    \includegraphics[width=0.8\linewidth]{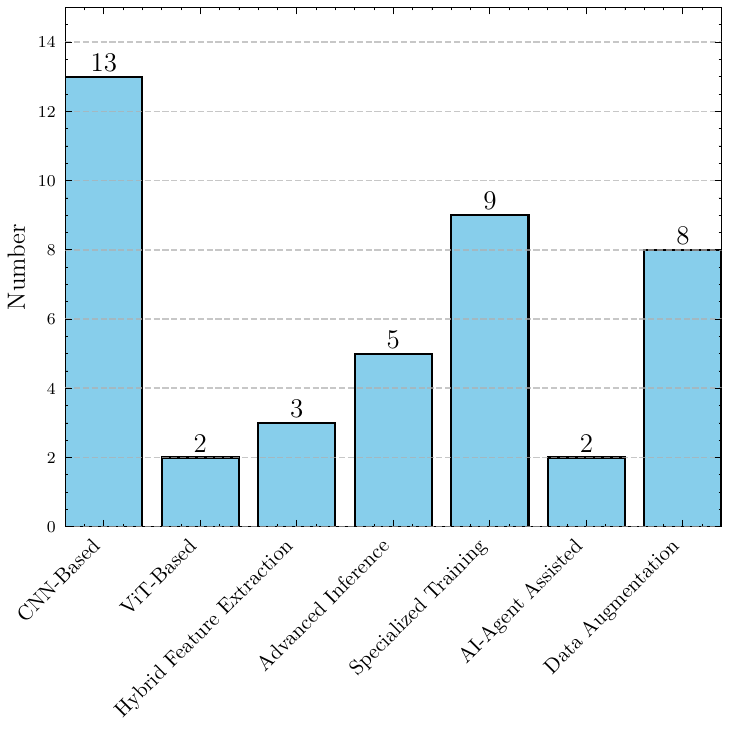}
    \caption{A summary of the number of submissions for the final evaluation in methodology categories.}
    \label{fig:methods}
    \vspace{-0.2cm}
\end{figure}

Figure~\ref{fig:methods} summarizes the final submitted methods for our final evaluation (see Sec.~\ref{sec:evaluation}) in different methodology categories. Overall, CNN-based architectures were the most common (13 submissions), followed by specialized training strategies such as custom losses, post-hoc uncertainty calibration, denoising U-Net, reinforcement Learning, or robust filtering (9 submissions) and geometric or domain-specific data augmentation approaches (8 submissions). Advanced inference schemes beyond direct regression, including simulation-based inference, normalizing flows, and MCMC, appeared in 5 submissions. Hybrid pipelines combining deep learning with fixed mathematical or physics-based descriptors such as scattering transforms or handcrafted cosmology features appeared in 3 submissions, while Vision Transformer-based architectures and explicitly AI-agent-assisted pipelines each appeared in 2 submissions. These counts suggest that the strongest submissions combined architectural inductive bias, augmentation, and careful uncertainty handling rather than relying on architecture choice alone. We also note that the best score achieved by a higher-order-statistics-based method on the public leaderboard is $9.1654$, indicating that under the current benchmark design, ML or hybrid learned approaches generally outperformed traditional summary-statistic-based methods.

\end{document}